\documentclass[preprint]{elsarticle}

\pdfoutput=1

\usepackage{graphicx}
\usepackage{color}
\usepackage{colortbl}
\usepackage{bbm}
\usepackage{amsfonts,amssymb,amsmath}
\usepackage[ruled,vlined]{algorithm2e}
\usepackage{url}
\usepackage{stmaryrd}
\usepackage{ulem}
\usepackage{hyperref}
\usepackage{breakurl}
\usepackage{algorithmic}

\def\foorp{{\hfill$\square$}}

\def\inv{{^{-1}}}

\def\RR{{\mathbb R}}

\def\EE{{\mathbb E}}

\def\Ex#1{{\EE\left\{#1\right\}}}

\def\CC{\mathbb C}
\def\ZZ{\mathbb Z}

\def\be{\begin{equation}}
\def\beq#1{\begin{equation}\label{#1}}
\def\ee{\end{equation}}
\def\bea{\begin{eqnarray}}
\def\beqa#1{\begin{eqnarray}\label{#1}}
\def\eea{\end{eqnarray}}
\def\ba{\begin{array}}
\def\ea{\end{array}}

\DeclareMathAlphabet{\mathpzc}{OT1}{pzc}{m}{it}

\def\cC{{\mathcal C}}

\def\cG{{\mathcal G}}

\def\cL{{\mathcal L}}

\def\cS{{\mathcal S}}

\def\cW{{\mathcal W}}

\def\Lone{L^1(\RR )}
\def\Ltwo{L^2(\RR )}

\def\ltwo{\ell^2(\ZZ )}

\def\bds{_{-\infty}^\infty}

\def\bpsi{\boldsymbol{\psi}}

\def\ba{{\mathbf a}}

\def\bg{{\mathbf g}}
\def\bG{{\mathbf G}}
\def\bh{{\mathbf h}}

\def\bN{{\mathbf N}}

\def\bU{{\mathbf U}}

\def\bx{{\mathbf x}}
\def\bX{{\mathbf X}}
\def\by{{\mathbf y}}
\def\bY{{\mathbf Y}}
\def\bZ{{\mathbf Z}}

\def\v{{\underline{v}}}

\newcommand{\argmin}{\mathop{\mathrm{argmin}}}

\newtheorem{theorem}{Theorem}
\newtheorem{proposition}{Proposition}
\newtheorem{lemma}{Lemma}

\newtheorem{remark}{Remark}

\def\bpsi{{\boldsymbol\psi}}

\renewcommand\bg{{\boldsymbol g}}
\renewcommand\bG{{\boldsymbol G}}
\renewcommand\bx{{\boldsymbol x}}
\renewcommand\by{{\boldsymbol y}}
\renewcommand\bN{{\boldsymbol N}}
\renewcommand\bU{{\boldsymbol U}}
\renewcommand\bX{{\boldsymbol X}}
\renewcommand\bY{{\boldsymbol Y}}
\renewcommand\bZ{{\boldsymbol Z}}
\def\bz{{\boldsymbol z}}
\def\bW{{\boldsymbol W}}

\def\cbG{{\boldsymbol{\bG}}}

\def\libr{\llbracket}
\def\ribr{\rrbracket}

\definecolor{forestgreen}{rgb}{0.13,0.54,0.13}
\definecolor{darkviolet}{rgb}{0.58,0,0.83} 

\journal{Applied and Computational Harmonic Analysis}

\begin{document}
\begin{frontmatter}

\title{Time-frequency and time-scale analysis of deformed stationary processes, with application to non-stationary sound modeling}
\author{H. Omer}
\author{B. Torr\'esani\corref{cor1}}
\ead{bruno.torresani@univ-amu.fr}
\cortext[cor1]{Corresponding author; Phone: +33 4 13 55 14 12; Fax: +33 4 13 55 14 02}
\address{Aix-Marseille Universit\'e,  CNRS, Centrale Marseille, I2M, UMR 7373, 13453 Marseille, France}

\date{\today}


\begin{abstract}
A class of random non-stationary signals termed {\it timbre$\times$dynamics}
is introduced and studied. These signals are obtained by non-linear transformations
of stationary random gaussian signals, in such a way that the transformation can be
approximated by translations in an appropriate representation domain. In such
situations, approximate maximum likelihood estimation techniques can be derived,
which yield simultaneous estimation of the transformation and the power spectrum
of the underlying stationary signal.

This paper focuses on the case of modulation and time warping of stationary signals,
and proposes and studies estimation algorithms (based on time-frequency and time-scale
representations respectively) for these quantities of interest.

The proposed approach is validated on numerical simulations on synthetic signals, and
examples on real life car engine sounds.
\end{abstract}


\end{frontmatter}


\section{Introduction}
\label{se:intro}
Very often, statistical signal processing approaches and algorithms rest on
stationarity assumptions placed on the signals of interest and/or on the
noise, stationarity being understood as some statistical form of translation
invariance. While in many signal processing problems of interest such stationarity
assumptions make perfect sense (at least within sufficiently small segments of
the signal of interest), this is not always the case. There are situations where
departure from stationarity carries essential information on the studied phenomena,
and in which one needs to measure non-stationarity accurately. This is an old
problem, that has been considered in some specific situations. Most often,
the reference stationary signals are modeled as (deterministic) sine waves or
sums of sine waves (as is the case in modulation theory), and the object of
interest takes the form of instantaneous frequency ou group delay (see
e.g.~\cite{Flandrin99time} and references therein). Recent developments in this
domain involve approaches such as Empirical Mode Decompositions
(see~\cite{Huang05hilbert} for a review) and various instances of
reassignment and synchrosqueezing methods~\cite{Auger13time}, up to
the shape function analysis which allows one to step away from the
sine wave model~\cite{Wu13instantaneous}. All these approaches assume
underlying stationarity in a deterministic setting (sine waves, or shape
functions) and do not involve stochastic modeling of stationarity. However,
there are several cases of interest where one cannot assume that the
underlying signal is a sine wave or a sum of sine waves, because it
originates from complex systems that involve fluctuations.

A good example of such situations is provided by the classical
{\it shape from texture} problem of image processing (see for
example~\cite{Malik97computing,Clerc02texture}), where a regular (stationary) texture
is projected on a curved surface, yielding images with non-stationary textures.
When the underlying texture is periodic, deterministic approaches can be
exploited, but this is no longer true when the underlying texture is random.
Other examples, which we shall be interested in in this paper, come from audio
signal processing, where non-stationary sounds can often be associated to
non-stationary motions. Think for example of an accelerating car engine sound.
Our ear is perfectly capable to detect regime changes from the sound, more precisely
from the non-stationarity of the sound. That example is particularly interesting
as in that case, the non-stationarity may be, in first approximation, modeled as
a clock change (or time warping), i.e. a periodic system whose inner clock varies
as a function of time. This example is the very motivation for the non-stationary
sound models we are studying in this paper, which we term {\it timbre$\times$dynamics}
models: models in which a stationary random signal (whose power spectrum is
interpreted as {\it timbre}) is modified by some nonlinear function, which encodes
the dynamics of some underlying systems.

Such models have already been considered by M. Clerc and S. Mallat in the
context of the shape from texture problem (see~\cite{Clerc02texture},
and~\cite{Clerc03estimating} for a more complete mathematical and statistical
analysis of the approach). One main contribution was the observation that the
non-linear transformation involved in the non-stationarity can in some situations
approximately translate into a transport in some appropriate representation domain
(time-scale, time-frequency). In addition, it was shown that characterizing
the transport in question leads to estimates for the deformation.

The current paper builds on this approach, exploiting a slightly different point
of view, namely explicit modeling of the underlying stationary signal as a
(possibly complex) Gaussian random signal. Under such assumptions, the deformed
signals are also complex Gaussian random signals, and the transport property in
an appropriate representation domain alluded to above can be made quantitative using
simple approximation techniques. This yields an approximate  Gaussian (or
complex Gaussian) model from which maximum likelihood estimation can be formulated.

More precisely, we consider two deformation models: modulations (which are
conveniently studied in the short time Fourier transform domain), and time
warping (studied in the wavelet domain). In both cases, we provide approximate
expressions for the transform of the corresponding deformed stationary process,
together with error estimates. We also provide sufficient conditions for the
invertibility of the covariance matrix of the so-obtained complex Gaussian models,
and propose estimation algorithms. The theoretical results are complemented by numerical results, both on synthesized signals and on real signal (car engine sounds), that show that the proposed approach is much more accurate and robust than simpler approaches based upon local frequency (or scale) averages of Gabor (or wavelet) transforms.  In comparison, the approach is also more accurate and robust than the algorithm of Clerc and Mallat, that doesn't fully exploit the stationarity of the underlying signal.

This paper is organized as follows. After this introduction, we briefly
account in Section~\ref{se:NotBack} for the mathematical background this
work rests upon, and introduce our notations. Section~\ref{se:modul} is
devoted to the study of stationary random signals deformed by modulations,
presented in the finite-dimensional, discrete case and Section~\ref{se:warp}
develops the case of time warping. Numerical results are provided and
discussed in Section~\ref{se:NumRes}, before the conclusion. More technical
proofs are given in Section~\ref{se:proofs}.

Part of this paper (namely the frequency modulation estimation) is an
elaborated version, of a short paper published in a conference
proceedings~\cite{Omer13estimation} (where results were announced
without proofs). The original contributions of the present paper consist in
the proofs of the results of~\cite{Omer13estimation} (in the finite-dimensional case), a detailed treatment of the time warping estimation (which is presented in the continuous time setting, and requires slightly more sophisticated mathematical analysis), and extended numerical results, including validation on real signals.

\section{Notations and background}
\label{se:NotBack}
We give in this section a short account of the mathematical tools we shall be
using in the sequel. We shall treat both discrete time and continuous time
signals, and to avoid inconsistencies use the notation $\bx: t\to x(t)$ for
continuous time signals and $\bx: t\to x[t]$ in the discrete time case (functions and vectors are denoted by a bold symbol while their point values or components are denoted using a normal font).

\subsection{Random signal models}
We shall mostly be concerned with stationary stochastic signal models, which we
emphasize in two different settings. A discrete, finite-dimensional setting, and
a continuous-time, infinite-dimensional setting. This choice is motivated by the
transforms we use, which turn out to take simpler forms in these settings.

\subsubsection{Finite-dimensional models}
\label{sub:FinDimMod}
We start with the simplest case, and consider real or complex random signals
$\bX=\{X[0],\dots X[L-1]\}$ of length $L$, and their infinite length extension
obtained using periodic boundary conditions. We denote by $I$ the integer
interval $I=\libr 0,L-1\ribr$, and by $I^c$ the centered version obtained by
$I^c=\ribr -L/2,L/2\ribr$ for $L$ even and the analogous expression for $L$ odd.
Given such a random signal, we  denote by $C_\bX$ and $R_\bX$ respectively the
corresponding covariance and relation matrices, defined for
$m,n\in\libr 0, L-1\ribr$ by
\begin{eqnarray}
C_\bX[m,n]&=&\Ex{X_m\overline{X}_m}\ ,\\
R_\bX[m,n]&=&\Ex{X_mX_m}\ .
\end{eqnarray}

We shall be concerned with zero-mean (real) Gaussian and complex Gaussian signals.
Zero-mean Gaussian signals are characterized by their covariance matrix.
Zero-mean complex Gaussian signals are characterized by their covariance and
relation matrices. A complex Gaussian centered signal is termed circular if its relation
matrix vanishes (see e.g.~\cite{Picinbono94circularity}).

Given a real signal $\bX\in\RR^L$, the corresponding analytic signal
$\bZ\in\CC^L$ is obtained by zeroing the negative frequencies of the discrete
Fourier (DFT) transform of $\bX$ and doubling it on positive frequencies. More precisely, $\bZ$ is defined by its DFT as $\hat Z[k] = 0$ for $k\ge L/2$ and $\hat Z[k]=2\hat X[k]$ otherwise. If $\bX$ is a
Gaussian, wide sense stationary random signal with power spectrum $\cS_\bX$, it
is easily shown (see~\cite{Picinbono94circularity} for a proof in the continuous
time setting, which can be adapted {\it mutatis mutandis}) that $\bZ$ is a
circular complex Gaussian wide sense stationary signal, therefore characterized
by its covariance matrix $C_\bZ[t,s] = \sum_{k\in I^+} \cS_X[k] e^{2i\pi k[t-s]/L}$.

\subsubsection{Continuous time models}
Because we are also interested in time warping as a signal deformation model, we
are naturally led to consider continuous time signal models, which we find
convenient to address also in the infinite support case. Since we are interested
in trajectories of random signals, the $\Ltwo$ model is not appropriate, and
signals are therefore modeled as (random) distributions. Discussing mathematical
aspects of such random distributions is far beyond the scope of the present
paper, and we simply refer to~\cite{Unser13sparse} (in particular chapter 3) for
a sketch of the theory.

For the present paper, we limit ourselves to a simple class of random
distributions, termed filtered Gaussian white noise in~\cite{Unser13sparse}.
Notions such as the generalized mean can be defined (in a distributional sense),
and covariance and relation matrices are also well defined as (distributional)
kernels of linear operators.

It will be sufficient for us to note that in such a context, measurements of the
form $\langle \bX,\varphi\rangle$ are well defined random variables, provided
the measurement vectors belong to some appropriate test function space (here,
the Schwartz class $\cS(\RR)$). The main aspect of the approach we develop below
is based upon transforms such as Gabor or wavelet transforms, both of which can
be well defined using such a language.

\subsection{Linear time-frequency and time-scale transforms}

\subsubsection{Gabor transform}
We consider the finite-dimensional Gabor transform, as studied in
e.g.~\cite{Soendergaard07finite}. Given a window function $\bg\in\CC^L$, the
Gabor transform $\cG_\bx$ of $\bx\in\CC^L$ with window $\bg$, hop size $a$
and frequency step $b$ (both $a$ and $b$ must be divisors of $L$) is defined
for $m\in\libr 0, M-1\ribr$ and  $n\in\libr 0\dots N-1\ribr$ by
\beq{fo:DGT}
\cG_\bx[m,n] = \langle \bx,\bg_{mn}\rangle=\!
\sum_{t=0}^{L-1} x[t]e^{-2i\pi mb[t-na]/L}\overline{g}[t\!-\!na]
\ee
with $M=L/b$ and $N=L/a$ and where $\bg_{mn}$ denote Gabor atoms of the form
$\bg_{mn}(t)=e^{2i\pi mb[t-na]}g[t-na]$. $\cG_\bX$ is therefore an $M\times N$ array.
For suitably chosen $\bg$, and $a$ and $b$ small
enough, the Gabor transform is invertible
(see~\cite{Carmona98practical,Grochenig01foundations});
in finite dimensional situations, efficient algorithms have been developed and
implemented (see~\cite{ltfatnote011}).

\begin{remark}
\label{rem:semicont.dgt}\rm
Despite the fact that we are working here in a discrete time setting, notice that the Gabor transform can also be defined for any value of the frequency variable. We shall denote by $\cG_\bx(.,.)$ (with parentheses instead of brackets) the corresponding semi-continuous Gabor transform, defined as
\beq{fo:DGT.semicont}
\cG_\bx(\nu,n) =
\sum_{t=0}^{L-1} x[t]e^{-2i\pi \nu[t-na]/L}\overline{g}[t\!-\!na]
\ .
\ee
\end{remark}
A simple way to obtain a local frequency estimation using Gabor transform is provided by the {\it local frequency}, which we define as the function
\beq{fo:loc.freq}
n\to \nu[n] = \frac{\sum_m mb |\cG_\bx[m,n]|^2}{\sum_m |\cG_\bx[m,n]|^2}\ ,
\ee
i.e. the average value of the frequency variable, using the
(suitably renormalized) spectrogram as probability density function.
We shall use that quantity as baseline method for the sake of comparison.
\subsubsection{Continuous wavelet transform and discretization}
Wavelet transform is based upon the idea of dilation, which is intrinsically a
continuous-time concept. Given the context of our work, and the fact that we are
interested in arbitrary dilation factors, we are naturally led to continuous
time models and the corresponding continuous wavelet transform (CWT for short),
constructed as follows. Let $\bpsi\in\Ltwo$. The corresponding continuous wavelet
transform of $\bx\in\Ltwo$ is the function $\cW_\bx$ defined as
\beq{fo:CWT}
\cW_\bx(s,u)\!=\!e^{-s/2}\!\!\int\!\! x(t)\overline{\psi}\left(e^{-s}(t-u)\right)dt
\!=\!\langle \bx,\bpsi_{(s,u)}\rangle\, ,
\ee
where $\bpsi_{(s,u)}$, defined by $\psi_{(s,u)}(t) = e^{-s/2}\psi(e^{-s}(t-u))$
is a shifted and rescaled copy of the wavelet $\bpsi$. Here, $s$ and $u$ take values in $\RR$.
Under suitable assumptions on $\bpsi$ (the so-called admissibility
condition, see e.g.~\cite{Carmona98practical}), the continuous wavelet transform is
invertible, a property which we will not exploit here. However, under some mild
assumptions, the admissibility condition implies that $\hat\psi(0)=0$, therefore
the CWT may be viewed as a (continuous) filter bank involving band pass filters.

The CWT may be discretized in several different ways. We choose here the
so-called dyadic wavelet transform~\cite{Mallat89theory} (also called stationary,
or translation invariant, wavelet transform), which is obtained using a regular sampling of the log-scale variable $s=m\ln(q)$ and a regular, scale-independent sampling $u=na$ of the time variable. Given corresponding such sampling constants $q$ and
$a$, the corresponding wavelets and wavelet transform read
\begin{eqnarray}
\label{fo:discr.wavelet}
\psi_{mn}(t) &=& q^{-m/2}\psi(q^{-m}(t-na))\\
\label{fo:discr.wt}
\cW_\bx[m,n]&=&\cW_\bx(m\ln(q),na)=\langle \bx,\bpsi_{mn}\rangle\ .
\end{eqnarray}

The CWT is readily extended to the analysis of distributions, provided the
wavelet is chosen in a suitable test function spaces. Hereafter, we shall assume
that $\bpsi\in\cS(\RR)$, which will allow us to analyze tempered distributions.

As in the Gabor case, a simple wavelet based time-dependent scale
estimation can be obtained by the {\it local scale}, which we define
as the function
\beq{fo:loc.scale}
n\to \sigma[n] = \frac{\sum_m {q^m} |\cW_\bx[m,n]|^2}{\sum_m |\cW_\bx[m,n]|^2}\ .
\ee
Again, we shall use that quantity for the sake of comparison.

\section{Deformation by modulation and estimation}
\label{se:modul}
We first consider modulations, that may suitably be approximately described as
frequency shifts in a joint time-frequency domain (see
Theorem~\ref{th:TFModApprox} below). Results of these section have been given
(without proof) in~\cite{Omer13estimation}.
We formulate the problem as follows.
\subsection{Model and estimates}
We limit ourselves to finite-dimensional situations, and use the notations
of Section~\ref{sub:FinDimMod} (the infinite-dimensional case is developed in~\cite{Omer15Modeles}).
Let $\bX\in\RR^L$ be a zero-mean, wide sense stationary Gaussian random
process, with covariance matrix $C_\bX$, and let $\bZ\in\CC^L$ denote
the associated analytic signal.
Let $\cS_\bX$ be the power spectrum of $\bX$, and assume that $\cS_\bX[0]=0$,
and if $L$ is even, that $\cS_\bX[L/2]=0$. Following the lines
of~\cite{Picinbono94circularity}, it is easily shown that under such an
assumption, $\bZ$ is a circular complex Gaussian random vector, therefore
completely characterized by its covariance matrix. The latter reads
\begin{equation}
C_\bZ[t,s] = \sum_{\nu \in I^+} \cS_\bX[\nu] e^{2i\pi\nu[t-s]/L}\ ,
\end{equation}
and $\bZ$ is therefore wide-sense stationary.

For simplicity, we work with complex-valued signals.
The observation is assumed to be an USB (upper sideband)
modulated version $\bY$ of the reference stationary signal $\bX$, namely
\begin{equation}
\label{fo:signal.model}
Y[t] = Z[t] e^{2i\pi\gamma(t)/L} + N[t]\ ,
\end{equation}
where $\gamma\in C^2([0,L])$ is an unknown smooth, slowly varying {\it modulation
  function}, and $\bN=\{N[t],\,t\in I_L\}$ is a circular complex
Gaussian white noise, with variance $\sigma_0^2$.
\begin{remark}\rm
When $\gamma$ is an affine function, $\bY$ is still a stationary random process.
As the estimation procedure we shall be studying below is mainly based on a
stationarity assumption placed in $\bZ$ (in addition to a smoothness assumption
on $\gamma$), this implies that the modulation $\gamma$ can only be determined up
to an affine function.
\end{remark}
When $\gamma$ is not an affine function, $\bY$ is not wide sense
stationary any more. The problem at hand is to estimate the unknown
modulation $\gamma$ and the original power spectrum $\cS_\bX$ from a single
realization of $\bY$.

\medskip
We will base the frequency modulation estimation on a Gabor representation of the observed signal, and deliberately disregard correlations across time of the Gabor transform (hence focusing on time slices of the Gabor transform of the observation).
The distribution of time slices of the Gabor transform $\cG_{\bZ}$ of the
analytic signal $\bZ$ associated with the original signal $\bX$ is characterized
in the following result, which result from direct calculations (see for
example~\cite{Carmona98practical} for similar calculations)
\begin{proposition}
\begin{enumerate}
\item
For fixed $n$, the Gabor transform $\cG_{\bN}[.,n]$ of the Gaussian white noise
is a stationary Gaussian random vector, with circular covariance matrix
\begin{equation}
C_{\cG_\bN}[m,m'] = \sigma_0^2\sum_{k=0}^{L-1} \overline{\hat g}[k]\hat g[k-(m'-m)b]
\end{equation}
\item
For fixed time index $n$, the Gabor transform $\cG_{\bZ}[.,n]$ of the analytic
signal is a circular complex Gaussian random vector, with covariance matrix
\begin{equation}
C_{\cG_\bZ}[m,m'] = \sum_{k \in I^+} \cS_\bX[k] \overline{\hat g}[k-mb]\hat g[k-m'b]
\end{equation}
\end{enumerate}
\end{proposition}
The estimation of the modulation will be based upon an approximation of the
covariance matrix of the observed signal. In a few words, the Gabor transform of
the frequency modulated signal can be approximated by a deformed version of the
Gabor transform of the original signal. The deformation takes the form of a time-varying frequency shift, as follows from a first order Taylor expansion of the modulation function. Since the Gabor atom $\bg_{mn}$ is localized around $t=na$, we just write
$\gamma(t)\approx \gamma(na) + (t-na)\gamma'(na)$, and thus obtain
%
%
$$
\cG_\bZ[m,n]\approx e^{2i\pi\gamma(na)/L}\cG_\bZ(mb+\gamma'(na),na)\ .
$$
Notice that since $\gamma'(na)$ has no reason to be an integer multiple of the frequency step $b$, we are led to a version of the Gabor transform defined in the continuous-frequency case, thus the use of parentheses instead of brackets for the arguments of $\cG_\bY$ (see Remark~\ref{rem:semicont.dgt}).

A more precise argument, exploiting the smoothness of the modulation function $\gamma$, leads to the
following result. Given a time index $n$ and some number $\delta$, denote by
$\bG^{(\delta;n)}$ the corresponding frequency-shifted copy of Gabor transform
$\cG_\bY$:
\begin{eqnarray}
\nonumber
\bG^{(\delta;n)}[m] \!&\!=\!&\! e^{2i\pi\gamma(na)/L}\sum_{t=0}^{L-1} Z_t
\overline{g}[t\!-\!na] e^{-2i\pi[m-\delta][t-an]/M }\\
&&\hphantom{aaaaaaaaaaa} +\ \cG_\bN[m,n] \ .
\label{eq:approximation2}
\end{eqnarray}
Let us set $\delta=\gamma'(na)/b$, and introduce the following constants
\begin{equation}
\mu_1 = \sum_{t\in I_1}|g[t]| \ , \; \mu_2 = \sum_{t \in I_2} t^2 |g[t]|\ ,\;
T = \sqrt{\frac{L}{\pi {\| \gamma'' \|}_{\infty}}}
\end{equation}
where  $I_2=\libr -T,T\ribr$ and $I_1=I^c\backslash I_2$.
Then we have

\smallskip
\begin{theorem}
\label{th:TFModApprox}
Assume that $L>4/\pi\|\gamma''\|_\infty$.
\begin{enumerate}
\item
For fixed time $n$, the Gabor transform $m\to \cG_\bY[m,n]$ may be approximated as
\begin{equation}
\cG_\bY[m,n]= \bG^{(\delta;n)}[m] + R[m]\ \label{eq:approximation1},
\end{equation}
and the remainder is bounded as follows: for all $m,m'$,
\begin{equation}
\left|\Ex{R[m]\overline{R}[m']}\right| \le
\sigma^2_Z \left( 2\mu_1 +\frac{\pi e}{L} \| \gamma''\|_{\infty} \mu_2 \right)^2\!\!,
\end{equation}
where  $\sigma_Z^2$ is the variance of $Z$.
\item
Given $\delta$, and for fixed $n$, $\bG^{(\delta;n)}$ is distributed
following a circular multivariate complex Gaussian law, with covariance matrix
\begin{equation}
C_{\bG^{(\delta;n)}}[m,m']=C_{\cG_\bZ}[m-\delta,m'-\delta]
\!+\! \ C_{\cG_\bN}[m,m']\ .
\label{eq:cov}
\end{equation}
\end{enumerate}
\end{theorem}
\underline{\it Proof:} The proof is given in Section~\ref{se:proofs}.\foorp

\medskip
The estimation procedure described below is a maximum likelihood
approach, which requires inverting the covariance matrix of
vectors $\bG^{(\delta;n)}$.
The latter is positive semi-definite by construction, but not necessarily
definite. The result below provides a
sufficient condition on $\bg$ and the noise for invertibility.
\begin{proposition}
\label{th:GabCovmatInv}
Assume that the window $\bg$ is such that
\begin{equation}
K_\bg:=\frac1{b}\min_{t\in\libr 0,M-1\ribr}\left(\sum_{k=0}^{b-1}|g[t+kM]|^2\right)>0\ .
\end{equation}
Then for all $\bx\in\CC^M$,
\begin{equation}
\langle C_{\bG}\bx,\bx\rangle \ge \sigma_0^2 K_\bg\,\|\bx\|^2\ ,
\end{equation}
and the covariance matrix is therefore boundedly invertible.
\end{proposition}
\underline{\it Proof:} The proof is given in the Section~\ref{se:proofs}.\foorp

\medskip
\begin{remark}
\label{rem:condition}\rm
The condition may seem at first sight unnatural in terms of Gabor frame theory. However,
it simply expresses that the number $M$ of frequency bins should not be too large
if one wants the covariance matrix to be invertible. Nevertheless, reducing $M$ also
reduces the precision of the estimate, and a trade-off has to be found, as
discussed in the next section.
\end{remark}

\subsection{Estimation algorithm}
\label{subsec:modul.estim}
We now describe in some details the estimation procedure corresponding
to our problem. The estimation problem is the following: from
a single realization of the signal model~\eqref{fo:signal.model}, estimate
the modulation function $\gamma$ and the original covariance matrix
$C_\bG$. We first notice the indeterminacy in the problem, namely the
fact that adding an affine function to $\gamma$ is equivalent to
shifting $\cS_\bX$. This has to be fixed by adding an extra constraint
in the estimation procedure.

The procedure is a two-step iterative approach: alternatively
estimate $\gamma$ given $C_\bG$, and estimate $C_\bG$
given $\gamma$.
\subsubsection{Maximum likelihood modulation estimation}
With the same notations as before, let $n$ be a fixed  value of the time index, and
let $\cbG=\cbG^{(n)}$ be the corresponding fixed time slice of $\cG_\bY$.
As the signal and therefore the fixed time Gabor transform slices are
distributed according to multivariate complex Gaussian laws,
the log-likelihood of a slice takes the form
\begin{equation}
\cL_\delta(\cbG) = -\left\langle C_{\bG^{(n;\delta)}}^{-1} \cbG,\cbG\right\rangle
- \ln\left(\pi^M \det(C_{\bG^{(n;\delta)}})\right)\ .
\end{equation}
Therefore,  maximum likelihood estimation leads to the frequency shift estimate
\begin{equation}
\hat\delta\! =\! \argmin_{\delta}
\left[\left\langle C_{\bG^{(\delta;n)}}^{-1} \cbG, \cbG\right\rangle +
\ln\left(\!\pi^M\!\! \det(\!C_{\bG^{(\delta;n)}}\!)\!\right)\!\right]\, .
\end{equation}
Notice that $\det(C_{\bG^{(\delta;n)}})$ actually does not depend on
the modulation parameter $\delta$. Hence the estimate
reduces to
\begin{equation}
\hat\delta = \argmin_{\delta} \left\langle C_{\bG^{(\delta;n)}}^{-1}
  \cbG,\cbG\right\rangle\ ,
\label{eq:ml1}
\end{equation}
a problem to be solved numerically. For the numerical solution, we resort to a simple exhaustive search, since the search domain is small. As $\delta(n)\approx  \gamma'(an)/b$, the estimates of $\delta$ for each $n$ lead to an estimate of $\gamma'$, which allows us to estimate $\gamma$ using standard integration and interpolation techniques. Notice that the estimation of $\gamma'$ is pointwise, no additional use of the smoothness assumption on $\gamma$ was necessary in our experiments.

Notice that this requires the knowledge of the covariance matrix $C_\bG^{(0;n)}$ corresponding to the Gabor transform of the noisy stationary signal. The latter is generally not available, and has to
be estimated as well.

\begin{remark}\rm
The minimisation problem~\eqref{eq:ml1} is solved by an exhaustive search on the $\delta$,
the estimate of $\gamma'$ is thus quantized. In addition, according to Remark~\ref{rem:condition},
the frequency step parameter $b$ has to be large for the invertibility of the covariance matrix,
and the quantization is therefore coarse. As described in~\cite{Omer13estimation}, this problem
is solved by computing a Gabor transform with fine frequency sampling, followed by a search on
a sequence of shifted coarsely frequency subsampled transforms. We refer to~\cite{Omer13estimation}
for more details, and to~\ref{rem:scale.subsampling} below where a similar procedure is described in the case of time warping estimation.

As a result, the quantization effect on the modulation function estimate
is significantly attenuated, as will be clear in the numerical simulations presented below in Section~\ref{se:NumRes}.
\end{remark}

\subsubsection{Covariance matrix estimation}
We now describe a method for estimating the covariance matrix $C_\bG^{(n;0)}$. 
Suppose that an estimate $\hat\gamma$ of the modulation function $\gamma$
is available. Then the signal $\bY$ can be demodulated by setting
\begin{equation}
\bU = \bY e^{-2i\pi\hat\gamma/L}\ , \label{eq:demodulation}
\end{equation}
Clearly, $\bU$ is an estimator of $\bZ  + \bN e^{-2i\pi\gamma/L}$, the noisy
stationary signal.
We can now compute the covariance matrix $C_{\cG_\bU}$ of the Gabor transform of
$\bU$, which is an estimator of $C_{\cG_\bZ} + C_{\cG_\bN}$. 
Comparing with equation~(\ref{eq:approximation2}) we finally obtain an 
estimator for the covariance matrix
\begin{equation}
C_{\cG_\bU} \approx C_{\bG^{(0;n)}}
\end{equation}

\subsubsection{Power spectrum estimation}
Suppose that an estimate $\hat\gamma$ of the modulation function $\gamma$
is available. Then the signal $\bY$ can be demodulated as in~\eqref{eq:demodulation},
and an estimate for the power spectrum deduced from it.
The power spectrum can be estimated using a standard Welch
periodogram estimator, or from a Gabor transform as described
in~\cite{Carmona98practical}.

\subsubsection{Summary of the estimation procedure}
We now summarize an iterative algorithm to jointly estimate the covariance
matrix $C_\bG^{(0;n)}$ and the modulation function $\gamma$, that exploits
alternatively the two procedures described above.
The procedure is as follows, given a first estimation of the modulation
function, we can perform a first estimation of the covariance matrix, which in
turn allows us obtain a new estimation of the modulation function.
The operation is repeated until the stopping criterion is satisfied.

For the initialization, we need a first modulation frequency estimate, for
which we use the local frequency function defined in~\eqref{fo:loc.freq}.

The stopping criterion is based upon the evolution of the frequency modulation
along the iterations. More precisely, we use the empirical criterion
\begin{equation}
\frac{||\hat\delta^{(k)} -\hat\delta^{(k+1)}||_2}{|| \hat\delta^{(k+1)}||_2} <
\epsilon, \label{eq:criterion}
\end{equation}
where $\delta^{(k)}$ is the estimation of $\delta$ at iteration number $k$.
The pseudo-code of the algorithm has been given in an earlier paper~\cite{Omer13estimation}.
We do not reproduce here as it is a mere modification of the time warping estimation algorithm to be described below.

\section{Time warping and estimation}
\label{se:warp}
We now turn to the second class of deformations that is of interest to us,
namely time warping. Again the assumption is that the observation is the image
of a stationary random signal by some transformation, here a time warping, that
breaks stationarity. To allow for arbitrary time changes, we have to consider
here a continuous time model. We consider smooth, monotonically increasing
{\it warping functions} $\gamma\in C^2$, and associate with them the corresponding
time warping operator $D_\gamma$, defined by
\beq{fo:time.warping}
[D_\gamma x](t)  =\sqrt{\gamma'(t)}\, x(\gamma(t))\ .
\ee
$D_\gamma$ is a unitary operator on $\Ltwo$ (i.e. $\|D_\gamma \bx\|=\|\bx\|$
for all $\bx\in\Ltwo$)
\begin{remark}\rm
Notice that additional assumptions are needed for $D_\gamma$ to be a well defined operator $\cS(\RR)\to\cS(\RR)$. A sufficient condition is that $\gamma \in \cC^\infty(\RR)$ be such that all its derivatives have at most polynomial growth, and satisfy the following condition 
\begin{equation*}
|\gamma(t)|>\alpha|t|^\beta\ ,\quad\forall t, |t| > t_0
\end{equation*}
for some $\alpha \in \RR^*_+$, $\beta \in \RR^*_+$ and $t_0\in\RR^+$.
\end{remark}
In addition, elementary calculus shows that
warping operators satisfy the composition property
$D_\gamma D_\varphi = D_{\varphi\circ\gamma}$ for all $\varphi,\gamma\in C^2$,
where the symbol $\circ$ denotes composition of functions. Similarly,
$(D_\gamma)\inv = D_{\gamma\inv}$, $\gamma\inv$ being the reciprocal function
of $\gamma$.

A warping operator $D_\gamma$ is said to be {\it controlled} if there exist two strictly positive constants $c_\gamma$ and $C_\gamma$ such that for all $t$,
\begin{equation}
\label{fo:control.warp}
0<c_\gamma\le \gamma'(t)\le C_\gamma<\infty_ .
\end{equation}
We shall need the following result, which follows from the unitarity of the
warping operator.
\begin{lemma}
\label{le:still.WN}
Let $\bN=\{N(t),t\in\RR\}$ be a (complex) Gaussian white noise. Then for all
monotonically increasing $\gamma\in C^2$, $D_\gamma\bN$ is also a (complex)
Gaussian white noise with the same variance.
\end{lemma}
\subsection{Model and estimates}
We therefore consider a generative model of the form
\beq{fo:time.warp}
Y(t) = [D_\gamma X](t) + N(t)\ ,
\ee
where $\bX$ is a second order stationary random process, $\bN$ is a white
noise with variance $\sigma_0^2$, and $\gamma$ is a smooth, monotonically
increasing function, such that the corresponding warping operator is controlled, as defined in~\eqref{fo:control.warp}. We denote again by $\cS_\bX$ the power spectrum of $\bX$.

Let $\bpsi\in\cS(\RR)$ be a wavelet function; we are interested in the behavior of
the discrete wavelet transform
\beq{fo:time.warp.WT}
\cW_\bY[m,n] = \langle \bY,\bpsi_{mn}\rangle
= \langle D_\gamma\bX,\bpsi_{mn}\rangle + \langle \bN,\bpsi_{mn}\rangle
\ ,
\ee
within a given region of the time-scale plane $m,n\in \Lambda$.
As remarked in~\cite{Clerc03estimating}, time warping can be approximated by
translation in the wavelet domain. 
In the framework of the Gaussian model, we
shall see that scale translations are enough to provide estimators for the time
warping.
\begin{remark}\rm
Clearly, when the warping function $\gamma$ is an affine function, the warped
signal $\bY$ is still second order stationary. Therefore, since our approach is
based upon non-stationarity, the function $\gamma$ can be determined only up to
an affine function.
\end{remark}

The behavior of the wavelet transform of stationary random signals is given by
\begin{proposition}
Let $\bpsi\in\cS(\RR)$ be an analytic wavelet function.
Let $\bX$ denote a wide sense stationary circular generalized Gaussian process with power spectrum
$\cS_\bX$, and let $\cW_\bX$ be its wavelet transform.
For any value of the time index $n$, the corresponding wavelet time
slice $\cW_\bX[,n]$ is a circular Gaussian random vector,
with covariance
$$
C_{\cW_\bX}[m,m'] =  q^{(m+m')/2} \langle \cS_\bX,  \hat\psi^{m}  \overline{\hat\psi^{m'}} \rangle,
$$
where $\hat\psi^{m}(\nu) ={\hat\psi}(q^m\nu)$. The latter can be formally written
$$
C_{\cW_\bX}[m,m'] = q^{(m+m')/2}\int_0^\infty\cS_\bX (\nu)
\overline{\hat\psi}(q^m\nu) \hat\psi(q^{m'}\nu) d\nu\ .
$$
\end{proposition}
In particular, if $\bX$ is a white noise with variance $\sigma_0^2$,
$$
C_{\cW_\bX}[m,m'] = \sigma_0^2 q^{(m+m')/2}\int_0^\infty
\overline{\hat\psi}(q^m\nu) \hat\psi(q^{m'}\nu) d\nu\ .
$$
We now turn to the analysis of time warped signals. As in the case of modulation, we shall see that if $\gamma$ is smooth, the effect of warping on wavelet transform can be approximated by in the time-scale plane, as shown in~\cite{Clerc03estimating}.
Assuming that the wavelet is localized around the origin $t=0$, and using a second order Taylor approximation for the warping function in the neighborhood of $\psi_{mn}$, namely $\gamma(t)\approx \gamma(na) + (t-na)\gamma'(na) + r(t)$ for
some remainder $r$, we obtain the approximation
$$
\cW_{D_\gamma\bX}[m,n] \approx \cW_\bX(m+\log_q(\gamma'(na),\gamma(na))\ .
$$
More precisely, let us set
\beq{fo:approx.wavelet.transform}
\bW^{(\delta;\tau)}[m]=\cW_\bX(m+\delta,\tau)\ + \cW_\bN(m,\tau)\ ,
\ee
which involves samples of a deformed version of the continuous
wavelet transform of $\bX$, and set $\delta_n= \log_q(\gamma'(na))$
and $\tau_n=\gamma'(na)$.
Notice that these samples need not (and will not in general) belong to
the sampling grid of the discrete wavelet transform. The following result, whose elementary proof is left to the reader, will be useful in the sequel.
\begin{lemma}
For fixed $n$, $m\to\bW^{(\delta_n;\tau_n)}[m]$ is a  circular Gaussian random vector with covariance matrix
$$
C_\bW^{(n)}[m,m']=C_{\cW_\bX}\! (m+\delta_{n},m'+\delta_{n})\\
+C_{\cW_\bN}[m,m']\ .
$$
\end{lemma}

\medskip
The approximation error can be controlled thanks to the following estimate
\begin{theorem}
\label{th:WTModApprox}
Let $\bpsi$ be a wavelet function satisfying
$|\psi(t)|\le (1+|t|^{\alpha})\inv$ with $\alpha>2$. Let $\bX$ be a stationary circular Gaussian generalized process,
whose power spectrum $\cS_\bX\in\Lone$ is such that
$$
\rho_\bX(\alpha):=
\int\bds|\nu|^{2-\frac{6}{\alpha+2}}\cS_\bX(\nu)d\nu<\infty\ .
$$
Let $\gamma\in C^2$ be a strictly increasing smooth function, satisfying~\eqref{fo:control.warp}, for some constants $c_\gamma,C_\gamma$, and let $\bY$ denote a noisy time-warped random signal, as defined in~\eqref{fo:time.warp}. Then
with the above notations, for each fixed time index $n$
the wavelet transform $\cW_\bY$ of the time-warped
signal $\bY$ may be approximated as
$$
\cW_\bY[m,n] = \bW^{(\delta_{n};\tau_n)}[m] + \cW_\bN[m,n] + \epsilon_{mn}\ ,
$$
where the remainder $\epsilon_{mn}$ is circular Gaussian random vector,
whose size can be bounded as
$$
\Ex{\!\left|\epsilon_{mn}\right|^2\!}\le \|\gamma''\|_\infty^2 q^{3m}\!
\left(\!K_0 \sigma_\bX\! +\! K_1 \|\gamma''\|_\infty^{\frac{-3}{\alpha+2}} q^{m}q^{\frac{-6m}{\alpha+2}}\sqrt{\rho_\bX\!(\alpha)}\!\right) ,
$$
where $K_0$ and $K_1$ are two constants that depend on $\alpha$.
\end{theorem}
\begin{remark}\rm
Similar estimates can be obtained starting from different decay assumptions on the analyzing wavelet $\psi$. We refer to~\cite{Omer15Modeles} for detailed derivations.
\end{remark}

\smallskip
\underline{\it Proof:}
The proof is mainly an adaptation of the proof of Theorem~\ref{th:TFModApprox}, that also takes into account decay assumptions on the wavelet, and is given in Section~\ref{prf:WTModApprox}.\foorp

\smallskip
We notice that for $\alpha$ large enough, the leading term of the error behaves as $q^{3m}$: the smoother the warping the better the estimate; also the error goes to 0 as the scale decreases.
However, let us stress that the estimation algorithm is not based upon such
an asymptotic analysis, but on an explicit modeling.

The estimate also seems to indicate (although the bound is not so easy to interpret) that large values of $\alpha$ should be preferred (i.e. well time localized wavelets). This is in accordance with the initial assumptions (i.e. $\psi\in\cS(\RR)$), and seems to be confirmed by the numerical simulations presented below.

\medskip

As before, the estimation of the warping function from a realization of the
deformed signal will require inverting the covariance matrix of fixed $n$ wavelet
transform. The result below provides a sufficient condition for invertibility,
and makes use of the so-called Fourier-Mellin transform. The latter associates
with any analytic signal $\bz$ (i.e. any $\Ltwo$ function whose Fourier
transform  vanishes on the negative half line) the function $\underline{\bz}$
of the real variable $s$ as
$$
\underline{z}(s) = \int_0^\infty \hat z(\nu) \nu^{-2i\pi\nu s}\frac{d\nu}{\sqrt{\nu}}\ .
$$
The Fourier-Mellin is a unitary transform (i.e. $\|\underline{\bz}\|=\|\bz\|$),
and we have the following
\begin{proposition}
\label{th:WavCovmatInv}
With the same notations as before, assume that the analyzing wavelet $\bpsi$ is
such that its Fourier-Mellin transform $\underline{\bpsi}$ satisfies
$$
K_\bpsi := \frac1{\log(q)}\inf_{s\in[0,1/\ln(q)]} \sum_{\ell\in\ZZ} 
\left|\underline\psi\left(s+\frac{\ell}{\log(q)}\right)\right|^2 >0
$$
Then for all $\bx\in\ltwo$,
$$
\langle C_\bW\bx,\bx\rangle \ge K_\bpsi \sigma_0^2\ \|\bx\|^2\ ,
$$
and the covariance matrix $C_\bX$ is therefore boundedly invertible.
\end{proposition}
\underline{\it Proof:}
the proof follows the line of the proof of Proposition~\ref{th:GabCovmatInv},
(moving from discrete to continuous, and replacing DFT by Mellin's transformation)
and is detailed in Section~\ref{prf:WavCovmatInv}.\foorp

Notice that the condition given in Proposition~\ref{th:WavCovmatInv} places restrictions on the wavelet and the scale variable discretization. When $q$ is too small, the periodized Fourier-Mellin transform of $\bpsi$ with period $1/\ln(q)$ can approach 0, and the covariance matrix may not be invertible. This will lead us to stick to large enough values of the scale sampling parameter $q$. Consequences are discussed in Remark~\ref{rem:scale.subsampling} below.

\subsection{Estimation algorithm}
The algorithm is essentially the same as the algorithm described in
Section~\ref{subsec:modul.estim}: it disregards time correlations, and searches
for a logarithmic scale translation for each fixed-time slice of a wavelet transform.

\subsubsection{Maximum likelihood modulation estimation}
Let $n$ be a fixed  value of the time index, and
let $\bW=\cW[\cdot,n]$ be the corresponding fixed-time slice of $\cW_\bY$ .

With the same assumptions as in Section~\ref{subsec:modul.estim} the maximum likelihood estimation for the scale shift takes the form
\begin{equation}
\hat\delta = \argmin_{\delta} \left\langle C_{\bW^{(\delta;\tau_n)}}^{-1}
  \bW,\bW\right\rangle\ ,
\label{eq:ml3}
\end{equation}
where $\delta \approx \log_q(\gamma'(na))$.
Notice that, as in the modulation problem, the knowledge of the covariance matrix $C_\bW^{(0;\tau_n)}$ is required. The latter matrix has to be estimated.

\subsubsection{Covariance matrix estimation}
The method for estimating the covariance matrix $C_\bW^{(n;0)}$ is very close to what we described for the modulation estimation, and mainly differs in the way we invert the time warping on $\bY$. 
Suppose that an estimate $\hat\gamma$ of the warping function $\gamma$ is available, $\bY$ can be time-unwarped by setting
\begin{equation}
\cW_{D_{\gamma^{-1}}\bY}[m,n] = \langle \bY,D_{\gamma}\bpsi_{mn}\rangle\ .
\label{eq:unwarping}                     
\end{equation}
From this we can estimate the needed covariance matrix using a standard sample estimator.

\subsubsection{Summary of the estimation procedure}
We summarize an iterative algorithm to jointly estimate the covariance
matrix $C_\bW^{(0;\tau_n)}$ and the warping function $\gamma$.
Algorithm is substantially the same as in Section~\ref{subsec:modul.estim}:
given a first estimation of the warping function, the covariance matrix is estimated using sample estimate from the unwarped signal as in~\eqref{eq:unwarping}; this in turn allows us obtain a new estimation of the warping function.
The operation is repeated until the stopping criterion~\eqref{eq:criterion} is satisfied.
For the initialization, we need a first warping function estimate, for
which we use the local scale function defined in~\eqref{fo:loc.scale}.

The pseudo-code of the algorithm can be found in Algorithm~\ref{al:WarpEst}.  

\begin{algorithm}
\caption{Joint covariance and warping function estimation}
\label{alg:estimation}

\begin{algorithmic}

\STATE {\bf Initialize} as in~\eqref{fo:loc.scale}
\WHILE{ criterion (\ref{eq:criterion}) is false}
\STATE $\bullet$ Compute $\hat\gamma^{(k)}$ by interpolation and integration from $\delta^{(k)}$.

\STATE $\bullet$ Unwarp $\bY$ using $\hat\gamma^{(k)}$
following~\eqref{eq:unwarping}

\STATE $\bullet$ Compute the wavelet transform of the unwarped signal
${\hat\bW}^{(k)}[m] = \cW^{(k)}_{D_{\gamma^{-1}}\bY}[m,n]$
\STATE $\bullet$ Given the covariance matrix of ${\hat\bW}^{(k;n)}$, estimate
$\hat{\delta}^{(k+1)}$ using~\eqref{eq:ml3}.

\STATE $\bullet$ $k:= k+1$

\ENDWHILE

\end{algorithmic}
\label{al:WarpEst}
\end{algorithm}

\begin{remark}
\label{rem:scale.subsampling}\rm
As a consequence of Proposition~\ref{th:WavCovmatInv}, the scale sampling step must be large enough to ensure the invertibility of the wavelet covariance matrix $\bW^{(0;\tau_n)}$. The optimization being performed using exhaustive search, this results in a quantization effect that can be quite significant. To reduce such an effect, we use the following procedure. The wavelet transform is computed on a fine scale sampling grid (i.e. with a small value of the scale sampling step $q$), but the optimization~\eqref{eq:ml3}
is performed independently on sub-grids of the form $Q_i=\{q_k = q^{i+kj},k=k_{min}\dots k_{max}\}$,
where $i$ runs from 0 to $j-1$ and $j$ is large enough to insure invertibility for the covariance matrix. Then the $\delta$ estimate from the best subgrid (i.e. yielding the maximal likelihood value) is retained. This procedure can be seen as a regularization that ignores correlations between close scales.
\end{remark}

\section{Numerical Results}
\label{se:NumRes}
We illustrate in this section the results obtained from the proposed approaches
in both cases, namely estimation of modulation or time warping. In the modulation case, we limit to results on synthetic signals for simplicity. We notice however that this case could be of practical interest, for instance for micro-Doppler estimation, of BLU demodulation. For the case of time warping,
numerical results include both simulations and results on real signals (car
engine sounds).

\subsection{Estimation of modulation}
The algorithm has been implemented using the Matlab scientific environment,
and the time-frequency tools implemented in the toolbox
{\sc Ltfat}~\cite{Soendergaard12linear}. Since corresponding numerical results
have already been published elsewhere, we simply illustrate here the method
on a synthetic signal, and refer to~\cite{Omer13estimation} for more details.

\begin{center}
\begin{figure}[!!h]
\centerline{
\includegraphics[width=12cm]{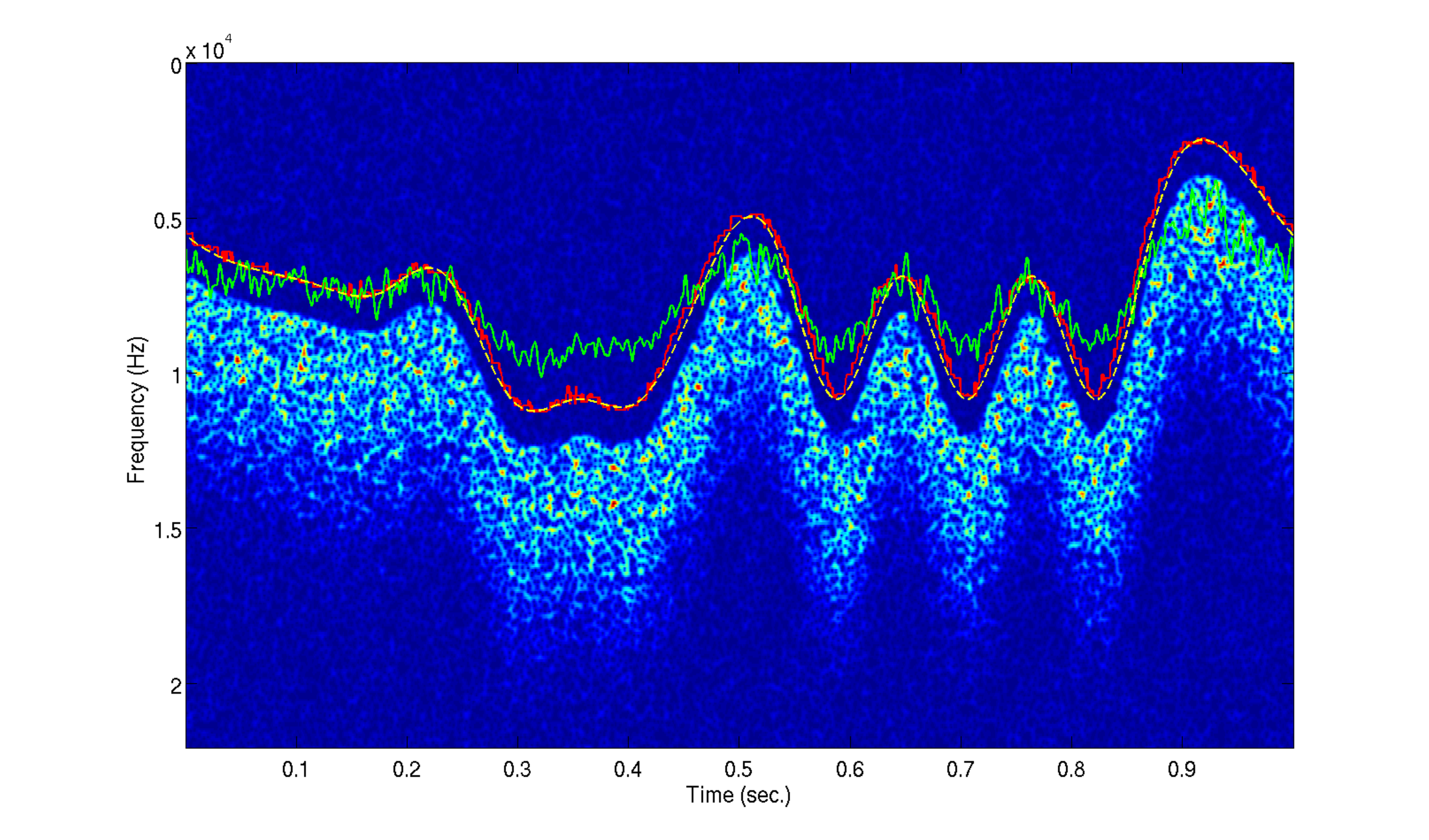}
}

\caption{
Gabor transform of a synthetic signal and frequency modulation
  estimation: ground truth (yellow, dotted) and estimates using the local
frequency function (green) and the proposed algorithm (red).}
\label{fig:modul.est}
\end{figure}
\end{center}

We display in Fig.~\ref{fig:modul.est} the results obtained using the
proposed algorithm on a synthetic random signal.
The original stationary signal $\bX$ was a band-pass filtered Gaussian white
noise, and the synthesized signal was obtained by frequency modulation with
a given  modulation function (displayed with a yellow curve on the
time-frequency plot). Gabor transform was computed using the {\sc Ltfat}
toolbox~\cite{Soendergaard12linear}, using a periodized Gaussian window
function. The squared modulus of the Gabor transform is displayed in
{\sc Fig}~\ref{fig:modul.est}, and the original frequency modulation (yellow)
and the estimated one (red) are superimposed. The algorithm is clearly able
to reproduce the frequency modulation curve with good accuracy.

We refer to~\cite{Omer13estimation} for more complete results. Since we
aim at studying real signals, for which the modulation model is often
not suitable, we limit the illustrations of the latter to that
example, and focus on the time warping model for which we present
more complete results below.

\subsection{Estimation of time warping}
\begin{center}
\begin{figure}[!!h]
\centerline{
\includegraphics[width=12cm]{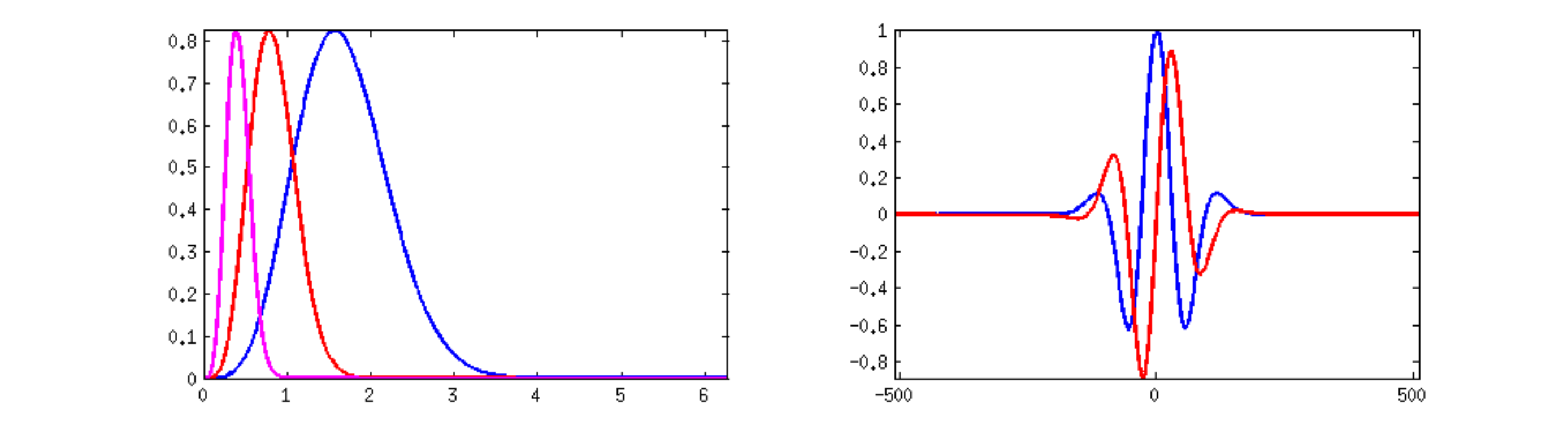}
}
\caption{Analytic derivative of Gaussian wavelets ($k=4$). Left: Fourier transforms, with three different values of the scale parameter. Right: real part (blue) and imaginary part (red) of a wavelet in the time domain.
}
\label{fig:wavelets}
\end{figure}
\end{center}

We now illustrate results obtained with the time warping estimation
algorithm, starting with the validation of the method on synthetic
signals. A synthetic time warped signal was generated, by warping a
band pass random stationary signal using a prescribed warping function.
The warping function derivative $\gamma'$ was then estimated (up to a
constant function) using the proposed method, as well as the local scale
function defined in~\eqref{fo:loc.scale} (which was also used for initializing
the proposed approach).
The wavelet transform was computed using an analytic derivative of Gaussian wavelet,
defined by its Fourier transform $\hat\bpsi$ which vanishes for negative frequencies
and reads on the positive half axis
$$
\hat\psi(\nu) = \nu^k e^{-\alpha\nu^2}\ \hbox{for}\ \nu\in\RR^+\ ,\qquad
\hat\psi(\nu) =0\ \hbox{for}\ \nu\in\RR^-
$$
with $\alpha$ a positive integer number, tuned so that the mode of $\hat\bpsi$ is located
at $\mathsf{fs}/4$, $\mathsf{fs}$ being the sampling frequency. Examples of such wavelets are displayed in Fig~\ref{fig:wavelets}. We notice that the parameter $\alpha$ controls the smoothness of
$\hat\bpsi$ at the origin of frequencies, and thus the decay of $\bpsi$, which play an
important role in the error estimates in Theorem~\ref{th:WTModApprox}.
The parameters of the wavelet analysis were set as follows:
we used a very fine scale sampling, $q=2^{-70}$, and $m$ was chosen the interval $m\in\libr 0;279\ribr$.

\begin{center}
\begin{figure}[!!h]
\centerline{
\includegraphics[width=12cm]{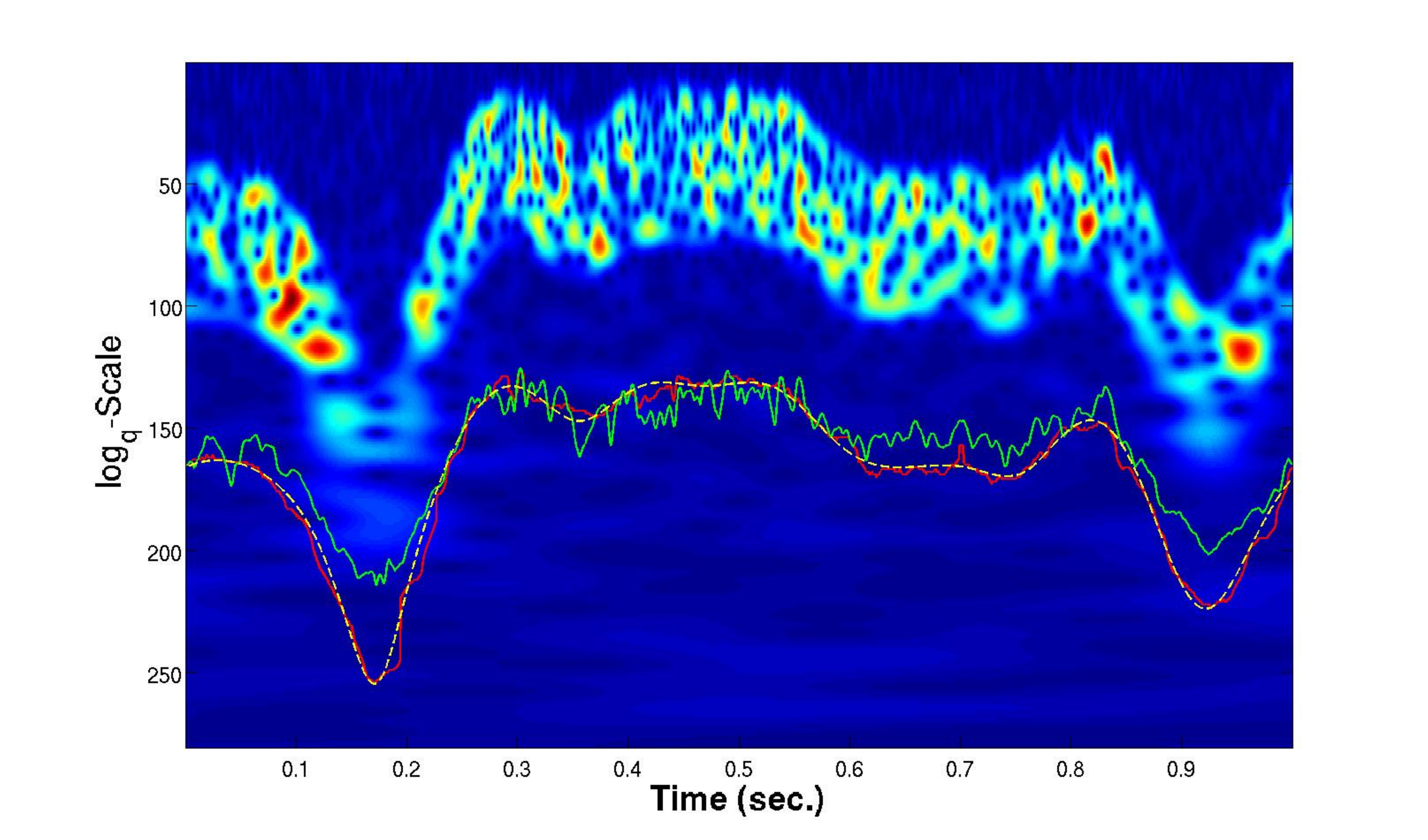}
}

\caption{Scalogram of a synthetic time warped random signal, and
warping function estimation: ground truth (yellow, dotted) and
estimates using the local scale function (green) and the
proposed algorithm (red).
}
\label{fig:modul.est2}
\end{figure}
\end{center}

We display in Fig~\ref{fig:modul.est2} the wavelet transform of a synthetic time warped stationary signal,
superimposed with the warping function derivative, together with two estimates: the local
scale function and the proposed method. As can be seen, similar to what was shown in the
case of modulation, the estimate provided by the proposed method turns out to be quite accurate.

To assess quantitatively the accuracy of the proposed algorithm and compare
with alternative approaches, we performed a simulation study: 500 realizations
of a time-warped random signal were generated following~\eqref{fo:time.warp},
using the same warping function, and the latter was estimated using the proposed
approach and the local scale function in~\eqref{fo:loc.scale}.  We provide in
{\sc Table}~\ref{tab:performances} the normalized average error (i.e. the euclidean norm of $\gamma'-\tilde\gamma'$, normalized by the euclidean norm of $\gamma'$, where $\tilde\gamma'$ is the estimate), averaged over all
realizations. We also provide in the same table the variances of the estimate.
The time-warping functions under consideration were sine waves with period
equal to the signal length (simulation 1), half the signal length
(simulation 2) and one fourth of the signal length (simulation 3).
The numerical results clearly show that the proposed approach
outperforms very significantly the results obtained using local scale, both in terms of error and variance. Three different wavelets were tested, with respectively $\alpha=70$, $\alpha=25$ and $\alpha=7$. Numerical results show that better localized wavelets (here $\alpha=70$) yield more accurate estimates, which is in accordance with Theorem~\ref{th:WTModApprox}. We didn't reproduce here results obtained with the Mallat \& Clerc algorithm, because the latter requires averagin over a large number of realizations of the random signal to yield results comparable to the ones described here. This originates from the fact that the estimate obtained in~\cite{Clerc03estimating} is very local and therefore sensitive to fluctuations. In addition, the iterative nature of our approach and the corresponding covariance re-estimation at each iteration seems to better exploit the underlying stationarity.

\begin{table}
\begin{center}
\begin{tabular}{|l||c|c|c|c|c|c|}
\hline
&\multicolumn{2}{l|}{Simulation 1}&\multicolumn{2}{l|}{Simulation 2}&\multicolumn{2}{l|}{Simulation 3}\\
&err.&var.&err.&var.&err.&var.\\
\hline
\hline
$\alpha=70$: Local scale &0.394&32.08&0.388&32.05&0.391&32.04\\
\hline
$\alpha=70$: Proposed method &0.115&6.61&0.126&7.76&0.151&10.44\\
\hline
\hline
$\alpha=25$: Local scale &0.248&19.53&0.259&21.56&0.358&28.55\\
\hline
$\alpha=25$: Proposed method &0.129&7.91&0.165&11.25&0.257&20.54\\
\hline
\hline
$\alpha=7$: Local scale &0.296&33.47&0.294&34.29&0.315&37.90\\
\hline
$\alpha=7$: Proposed method &0.226&24,75&0.236&26.47&0.275&33.03\\
\hline
\end{tabular}
\end{center}

\caption{Performance analysis of time warping estimation methods for three different analyzing wavelets (analytic derivative of Gaussian function of respective degree $\alpha=70$, top, $\alpha=25$, middle and $\alpha=7$, bottom): average error and variance of the estimate.}
\label{tab:performances}
\end{table}

\begin{figure}[!!h]
\centerline{
\includegraphics[width=12cm]{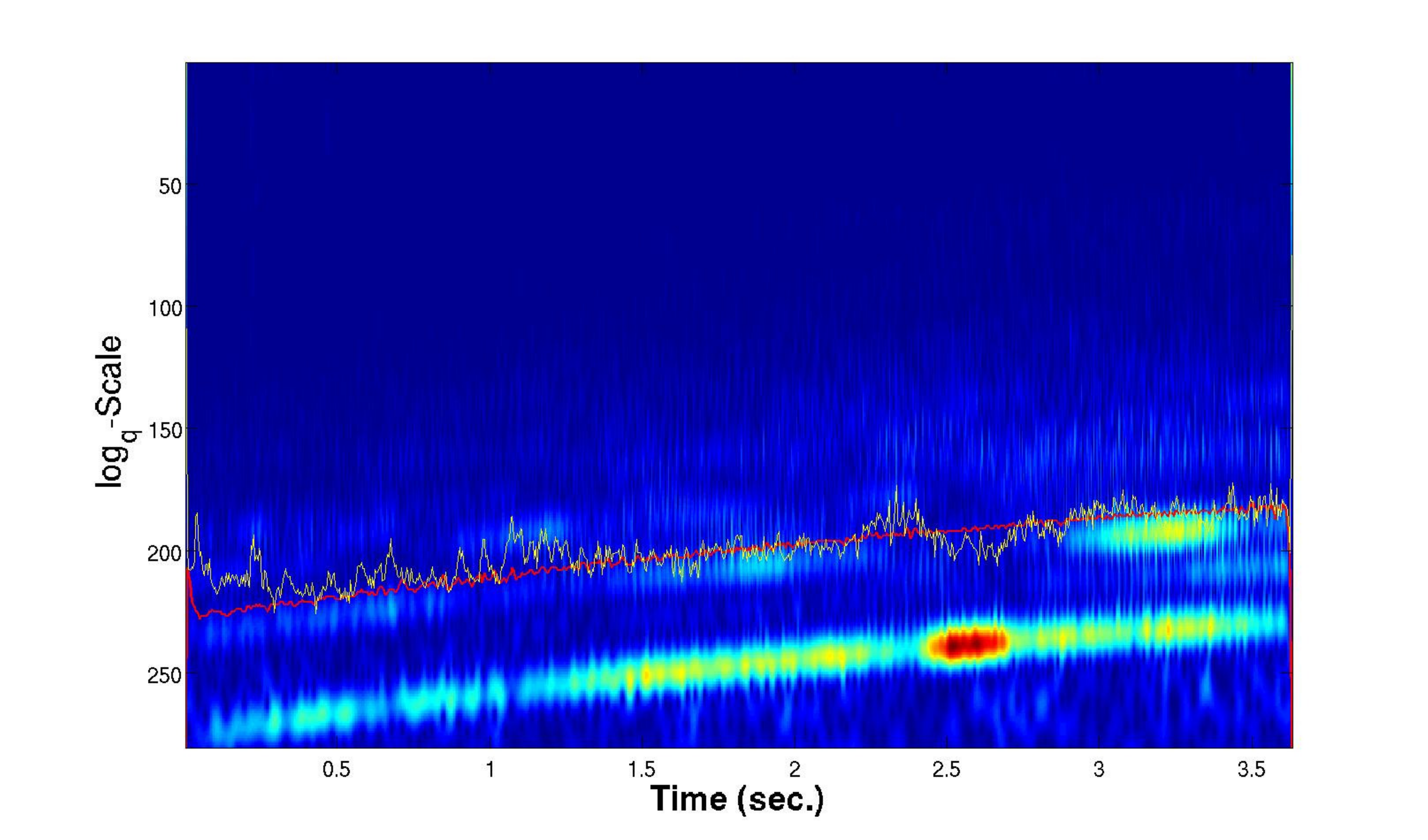}
}
\centerline{
\includegraphics[width=12cm]{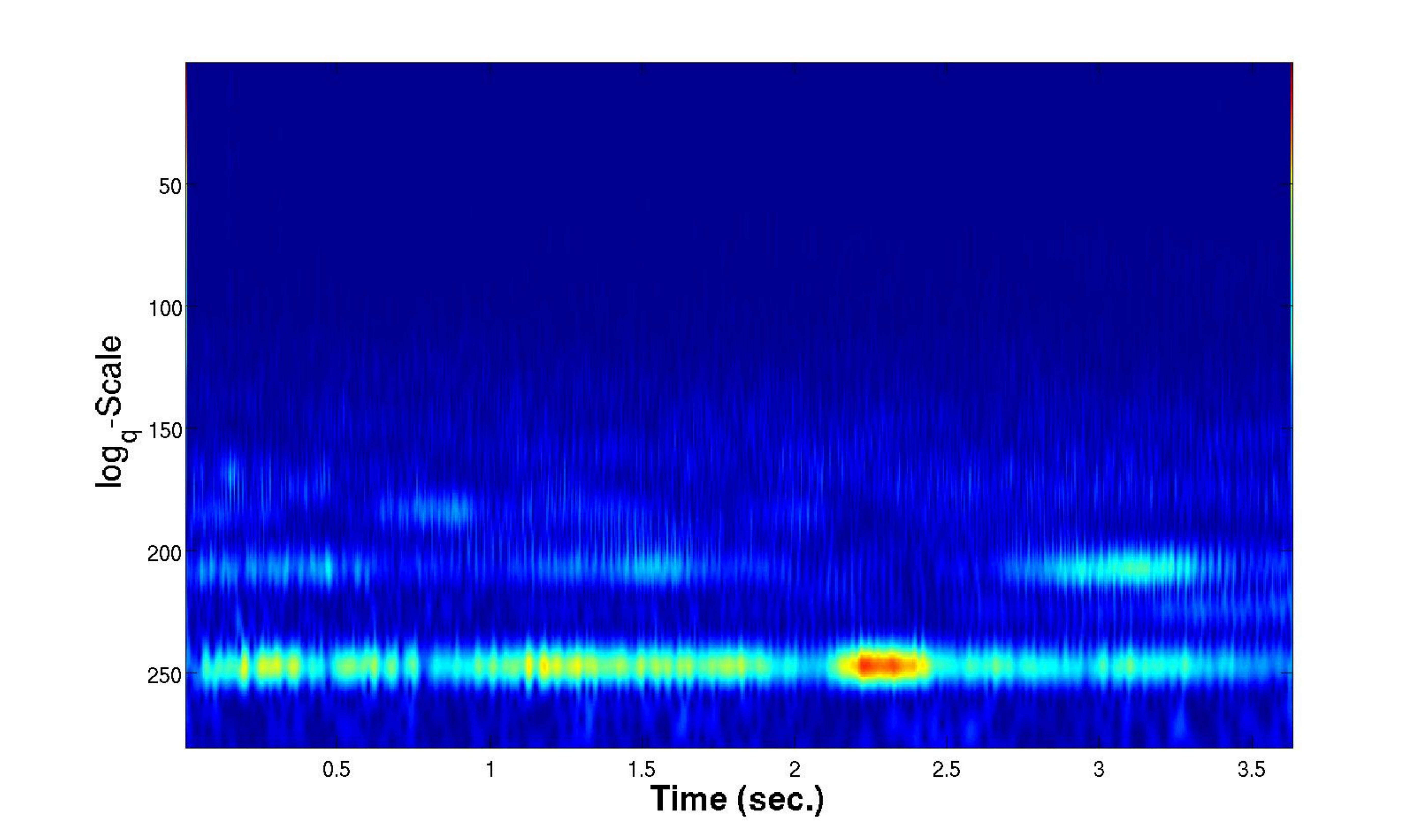}
}
\caption{Top: Scalogram of an accelerating car engine sound, and time
warping function estimation: estimates using the local scale function (yellow)
and the proposed algorithm (red).
Bottom: Scalogram of the "stationarized" signal.
}
\label{fig:golf}
\end{figure}

Finally, the algorithm was also tested on real signals. We display in
Fig.~\ref{fig:golf} (top) the result obtained on an accelerating car engine.
As before, the estimate provided by the proposed algorithm turns out to be
significantly smoother than the local scale function estimate. After estimation, the warping function can be inverted to produce a ``stationarized'' signal, the scalogram of which is also displayed in Fig.~\ref{fig:golf} (bottom).
Corresponding sound files are available (together with additional examples) at the web site

\centerline{\tt https://www.i2m.univ-amu.fr/$\sim$omer/SounDef/}

\noindent that show that the stationarized signal indeed captured the main aspects of the timbre of the original.

\section{Proofs of the main results}
\label{se:proofs}
\subsection{Proof of Theorem~\ref{th:TFModApprox}.}
The idea is to use the localization of the window function $g$ around the
origin, and ``freeze'' the modulation function $\gamma$ at the location $t=na$
where the Gabor function is centered.
For the sake of simplicity, let us set
\begin{eqnarray*}
G&=&\sum_{t=0}^{L-1} Z_t e^{2i\pi\gamma(t)/L}
\overline{g}[t-na] e^{-2i\pi  mb[t-na]/L}\ ,\\
\tilde G &=&\sum_{t=0}^{L-1} Z_t \overline{g}[t-na]
e^{2i\pi\gamma(na)/L} e^{-2i\pi [t-na][mb-\gamma'(na)]/L}\ ,
\end{eqnarray*}
and set $R=G-\tilde G$. Then
$$
R=\sum_{t=0}^{L-1} Z_t \overline{g}[t-na]e^{2i\pi\gamma(t)}e^{-2i\pi mb[t-na]/L}
[1-e^{2i\pi r(t)}]\ ,
$$
where $r(t)=\gamma(t)-\gamma(na)-[t-na]\gamma'(na)$; clearly,
$|r|\le\frac1{2}[t-na]\,\sup(|\gamma''|)$. Let us now consider the size estimate for
the remainder $\Ex{|R|^2}$. We consider the centered interval $I^c=\ribr-L/2,L/2\ribr$,
which we split into two parts, namely small values $I_1=\{t\in I^c: |t|\le T\}$
and large values $I_2=\{t\in I^c: |t|> T\}$, with
$T=\sqrt{L/\pi\|\gamma''\|_\infty}$. Then

\begin{eqnarray*}
\Ex{|R|^2}\! &\!=\!&\! \sigma_Z^2 \left|\sum_t \overline{g}[t-na]
e^{2i\pi\gamma(t)/L} \left[1-e^{2i\pi r(t)/L}\right]\right|^2\\
&\!\le\!&\!\!\! \sigma_Z^2\! \left( \sum |g[t]|
\left|1-e^{2i\pi r[t+na]}\right|\right)^2\\
&\!\le\!&\!\!\! \sigma_Z^2\! \left(2\!\sum_{t\in I_1} |g[t]|\! +\!
2\pi\!\! \sum_{t\in I_2}  |g[t]||r[t]|
e^{2\pi |r[t]|/L}\right)^2\\
&\!\le\!& \sigma_Z^2 \left(\! 2\mu_1 \!+\! \frac{\pi\|\gamma''\|_\infty}L\!
\sum_{t\in I_2}\! t^2 |g[t]| e^{\pi t^2\|\gamma''\|_\infty /L}\!\right)^2\\
&\le& \sigma_Z^2 \left( 2\mu_1 +
\frac{\pi e \|\gamma''\|_\infty}L \mu_2\right)^2\ ,
\end{eqnarray*}
where we have used the inequality $|e^z-1|\le |z| e^{|z|}$, and
where $\mu_1=\sum_{t\in I_1} |g[t]|$ and $\mu_2 =\sum_{t\in I_2} t^2|g[t]|$
characterize respectively the behavior of the window $\bg$ on its tail and near
the origin.

This concludes the proof.\foorp

\subsection{Proof of Proposition~\ref{th:GabCovmatInv}}
The covariance matrix of the Gabor transform (see for
example~\cite{Carmona98practical} for the general case) is given by
$$
C_\bG[m,m']=\sum_{\nu \in I_c} (\cS_\bZ[\nu-\delta]+\sigma_0^2)\overline{\hat g}[\nu-mb]\hat g[\nu-m'b]\ .
$$
Let $\bx\in\CC^M$, and compute
\begin{eqnarray*}
\langle C_\bG \bx,\bx\rangle &=& \!\!\!\!\sum_{m,m'=0}^{M-1}\!\!\!\!x[m]\overline x[m']
\sum_{\nu \in I_c}(\cS_\bZ[\nu-\delta]+\sigma_0^2) \hat g[\nu-m b]\overline{\hat g}[\nu-m'b]\\
&=&\sum_{\nu \in I_c}(\cS_\bZ[\nu-\delta]+\sigma_0^2) |\hat h[\nu]|^2\ge \sigma_0^2 \|\hat \bh\|^2\ ,
\end{eqnarray*}

where we have set $\hat h[\nu]=\sum_{m=0}^{M-1}\hat g[\nu-mb]x[m]$. Now,
$h[t]=(\sum_{m=0}^{M-1} x[m]e^{2i\pi mt/M})g[t]/L$ and the
standard periodization argument yields
\begin{eqnarray*}
\|\hat \bh\|^2 &=& L \|\bh\|^2\\
&=&\frac{L}{L^2}\sum_{t=0}^{L-1} |g[t]|^2
\left|\sum_{m=0}^{M-1} x[m]e^{2i\pi mt/M}\right|^2\\
&=&\frac{1}{L}\sum_{u=0}^{M-1} \sum_{k=0}^{b-1}|g[u-kM]|^2 
\left|\sum_{m=0}^{M-1} x[m]e^{2i\pi mu/M}\right|^2\\
&\ge& \frac{1}{L} \min_{u\in\libr0,M-1\ribr}\left( \sum_{k=0}^{b-1}|g[u-kM]|^2 \right) \|\hat \bx\|^2\\
&\ge& \frac{1}{b} \min_{u\in\libr0,M-1\ribr} \left( \sum_{k=0}^{b-1}|g[u-kM]|^2 \right) \|\bx\|^2\\
\end{eqnarray*}
This concludes the proof of the proposition.\foorp

It is worth noticing that this proof, as often in the case of Gabor
analysis, carries through {\it mutatis mutandis} to the case of
Gabor analysis for continuous time signals.

\subsection{Proof of Lemma~\ref{le:still.WN}.}
The proof follows from the unitarity of the warping operator.
The Gaussian white noise with variance $\sigma_0^2$ is characterized by its characteristic
function $P_\bN(\varphi) = \Ex{e^{i \langle\bN,\varphi\rangle}} = e^{-\|\varphi\|^2/2\sigma_0^2}$,
for all test function in the Schwartz space $\varphi\in\cS(\RR)$.
Similarly, the warped white noise's characteristic function reads
$$
P_{D_\gamma\bN}(\varphi) = \Ex{e^{i \langle D_\gamma\bN,\varphi\rangle}}
= \Ex{e^{i \langle \bN,D_\gamma\inv\varphi\rangle}}
=e^{-\|D_\gamma\inv\varphi\|^2/2\sigma_0^2}
=e^{-\|\varphi\|^2/2\sigma_0^2}
$$
which implies that $D_\gamma\bN$ is a Gaussian white noise, with the same variance.\foorp

\subsection{Proof of Theorem~\ref{th:WTModApprox}}
\label{prf:WTModApprox}
Let us compute the difference between the actual wavelet transform of the
deformed process and the approximation. Fix a time index $n$, and denote by
$\tilde\gamma_n(t)=\gamma(na)+(t-na)\gamma'(na)$ the first order approximation of
$\gamma$ in the neighborhood of the center $t=na$ of the wavelet $\psi_{mn}$. Set also
$D_\gamma^\circ = D_\gamma/\sqrt{\gamma'}$ and define $D_{\tilde\gamma_n}^\circ$
similarly. Write
\begin{eqnarray*}
\epsilon_{mn}&=&\cW_\bY[m,n] - \cW_\bX(m+\log_q(\gamma'(an)),\gamma(an))\\
&=&\left\langle(D_\gamma-D_{\tilde\gamma_n})\bX,\psi_{mn}\right\rangle
=\epsilon_{mn}^{(0)} + \epsilon_{mn}^{(1)}
\end{eqnarray*}
where we have set
\begin{eqnarray*}
\epsilon_{mn}^{(0)} &=& \left\langle\left(\sqrt{\gamma'}-\sqrt{\gamma'(na)}\right)D_\gamma^\circ \bX,\psi_{mn}\right\rangle\\
\epsilon_{mn}^{(1)} &=& \sqrt{\gamma'(na)}\left\langle(D_\gamma^\circ-D_{\tilde\gamma_n}^\circ)\bX,\psi_{mn}\right\rangle
\end{eqnarray*}

We bound those two terms independently. Consider first $\epsilon_{mn}^{(0)}$, and write
$$
\Ex{|\epsilon_{mn}^{(0)}|^2}=\int\bds\cS_\bX(\nu)\left|g(\nu)\right|^2d \nu\ ,
$$
where
\begin{eqnarray*}
g(\nu)&=&\left|\int\bds e^{2i\pi\nu\gamma(t)}
\left(\sqrt{\gamma'(t)}-\sqrt{\gamma'(na)}\right)\psi_{mn}(t)dt\right|\\
&\le &\frac{1}{2}\left\|\frac{\gamma''}{\sqrt{\gamma'}}\right\|_\infty q^{-m/2}\int\bds\!\!\!\!\! |t\!-\!na| 
\left|\psi\left(q^{-m}(t\!-\!na)\right)\right|dt\\
&\le & \frac{\mu}2\left\|\frac{\gamma''}{\sqrt{\gamma'}}\right\|_\infty q^{3m/2}\ ,\ \hbox{where }
\mu=\int\bds\!\!\!\!\!\! |s|\,|\psi(s)|\,ds\ .
\end{eqnarray*}
Assume now that $|\psi(t)|\le (1+|t|^\alpha)\inv$. Then
$$
\mu\le 2\int_0^\infty\!\!\!\frac{s\,ds}{1+|s|^\alpha}\le
2\int_0^1\!\! sds + 2\int_1^\infty\!\! s^{1-\alpha}ds\le \frac{\alpha}{2(\alpha-2)}
$$
All together, letting $\sigma_\bX^2=\|\cS_\bX\|_1$ be the variance of $\bX$, we obtain
$$
\Ex{|\epsilon_{mn}^{(0)}|^2}\le \frac{\alpha^2\sigma_\bX^2}{16(\alpha-2)^2}
\left\|\frac{\gamma''}{\sqrt{\gamma'}}\right\|_\infty^2 \,q^{3m}\ .
$$
\medskip
Consider now $\epsilon_{mn}^{(1)}=\sqrt{\gamma'(na)}
\left\langle(D_\gamma^\circ-D_{\tilde\gamma_n}^\circ)\bX,\psi_{mn}\right\rangle$. Then
\begin{eqnarray*}
\Ex{\left|\epsilon_{mn}^{(1)}\right|^2}\!\! &\!=\!&\!\! q^{-m}\gamma'(na) \int\bds\cS(\nu)\,\bigg|\int e^{2i\pi\nu\tilde\gamma_n(t)}\,\times\\
&&\!\!\left[1\!-\!e^{i\pi\nu(t-na)^2r(t)}\right]
\overline\psi\!\left((q^{-m}(t\!-\!na)\right)dt\bigg|^2\!\!\!d\nu\\
&\le& 4q^{-m}\gamma'(na)\int\bds\cS_\bX(\nu) f(\nu)^2d\nu\ ,
\end{eqnarray*}
where $f$ (that actually also depends on $n$ and $m$) is bounded as
\begin{eqnarray*}
f(\nu)\!\! &\!\!  = \!\! &\!\! \frac1{2}
\bigg|\int e^{2i\pi\nu\tilde\gamma_n(t)}\,\left[1\!-\!e^{i\pi\nu(t-na)^2r(t)}\right]
\overline\psi\!\left((q^{-m}(t\!-\!na)\right)dt\bigg|\\
&\!\!  \leq \!\! &\!\!\int\bds\! \left|\sin\left(\frac{\pi}2\nu(t\!-\!na)^2r(t)\right)\right|
|\psi\left(q^{-m}(t\!-\!na)\right)\!|dt\\
&\leq& q^{m}\int\bds|\psi(s)|\left|\sin\left(\frac{\pi}2\nu q^{2m}s^2r(t)\right)\right|ds\\
&\le& q^m\left[\int_{I_1(\nu)}\!\!\!\!  |\psi(s)|ds + \frac{\pi\|\gamma''\|_\infty}2|\nu| q^{2m}
\int_{I_2(\nu)}\!\!\!\!\!\!\! s^2|\psi(s)|ds\right] \ .
\end{eqnarray*}
Here we have introduced a splitting of
the real axis as the union of two domains (that depend on the frequency $\nu$)
$$
I_1(\nu) =\{s: |s| \geq u_0 \} ,\quad I_2(\nu)=\RR\backslash I_1(\nu)
$$
where $u_0$ is a free parameter that can be adapted.
We have also used the bounds $|\sin(u)|\le 1$ and $|\sin(u)|\le |u|$ within those two domains. 

Let us set for simplicity $C_0=\frac{\pi\|\gamma''\|_\infty}2|\nu| q^{2m}$.
Using the decay of the wavelet $\psi$, namely $|\psi(t)|\le (1+|t|^\alpha)\inv$, we obtain
$$
\int_{I_1(\nu)}|\psi(s)|ds\le 
\frac2{\alpha-1}u_0^{1-\alpha}\ ,\quad
\int_{I_2(\nu)}s^2|\psi(s)|ds\le\frac2{3}u_0^3\ .
$$
Then
$$
f(\nu)\le 2q^mF(u_0)\ ,
$$
where
$$
F(u_0)=\frac{u_0^{1-\alpha}}{\alpha-1} + 
 C_0 \frac{u_0^3}{3}\ .
$$
The value ${u_0}_{\sf opt}$ of $u_0$ that minimizes the latter expression is
easily found to be ${u_0}_{\sf opt} = C_0^{-\frac{1}{\alpha+2}}$,
which leads to the bound
$$
F({u_0}_{\sf opt})=\frac{\alpha+2}{3(\alpha-1)}C_0^{1-\frac{3}{\alpha+2}} .
$$
Therefore,
\begin{eqnarray*}
f(\nu)&\le &  2 q^m \frac{\alpha+2}{3(\alpha-1)} \left( \frac{\pi\|\gamma''\|_\infty}2|\nu| q^{2m}\right)^{1-\frac{3}{\alpha+2}} \\
\end{eqnarray*}
Putting things together, we finally obtain
\begin{eqnarray*}
\Ex{\left|\epsilon_{mn}^{(1)}\right|^2}&\le &
\frac{16(\alpha+2)^2}{9(\alpha-1)^2}\gamma'(an)\left(\frac{\pi\|\gamma''\|_\infty}{2}\right)^{2-\frac6{\alpha+2}}\\
&&\hphantom{??}\times \left(q^m\right)^{5-\frac{12}{\alpha+2}}
\int\bds|\nu|^{2-\frac{6}{\alpha+2}}\cS_\bX(\nu)d\nu\ .
\end{eqnarray*}
Finally,  bound the total error by 
\begin{eqnarray*}
\Ex{|\epsilon_{mn}|^2} \leq \sqrt{\Ex{|\epsilon^{(0)}_{mn}|^2}} + \sqrt{\Ex{|\epsilon^{(1)}_{mn}|^2}},
\end{eqnarray*} concludes the proof.\foorp

%
%
\subsection{Proof of Proposition~\ref{th:WavCovmatInv}.}
\label{prf:WavCovmatInv}
Let $n$ be a fixed time index. We consider the covariance matrix of the
approximate wavelet transform $\bW=\bW^{(\gamma(n);\log_q(\gamma'(n)))}$, which
takes the form
$$
C_\bW^{(n)}[m,m'] = \Ex{\bW[m]\overline{\bW}[m']}
 = q^{(m+m')/2} \left\langle (\cS_\bX + \sigma_0^2),  \hat\psi^{m}  \overline{\hat\psi^{m'}} \right\rangle .
$$
Let $\bx\in\ltwo$, and set 
$y(\nu)=\sum_{m=-\infty}^\infty q^{m/2}x[m]\hat\psi\left(q^m\nu\right)$,
for $\nu\in\RR^+$. Then
$$
\langle C_\bW^{(n)}\bx,\bx\rangle = \langle (\cS_\bX +\sigma_0^2),  {|y|}^2 \rangle
\ge \sigma_0^2 \|\by\|^2 \ge  \sigma_0^2 \|\underline{\by}\|^2\ ,
$$
with $\underline{\by}$ the Mellin transform of $\by$. An explicit
calculation shows that
$$
\underline{y}(s) =\underline{\psi}(s)\,
\hat x(s\ln(q))\ ,
$$
and therefore a standard periodization argument yields
\begin{eqnarray*}
\|\underline{\by}\|^2 &=& \int\bds |\underline{\psi}(s)|^2
|\hat x(s\ln(q))|^2\,ds\\
&=& \int_0^{1/\ln(q)} \!\!\!|\hat x(s\ln(q))|^2\sum_{\ell=-\infty}^\infty
\left|\underline{\psi}\left(s+\frac{\ell}{\ln(q)}\right)\right|^2
\,ds\\
&\ge &K_\bpsi \|\bx\|^2\ .
\end{eqnarray*}
Putting all estimates together yields the result.\foorp

\section{Discussion and conclusions}
\label{se:concl}
We have presented in this paper a new approach for the estimation of smooth deformations of Gaussian stationary random signals, that only exploit the stationarity and smoothness assumptions. Two types of deformations were considered, namely smooth modulation and smooth time warping, for which the transformation approximately manifests itself by translations in a suitable representation space. In both cases, an iterative algorithm was proposed, that estimates alternately the deformation and the covariance matrix of the signal. The analysis was very similar in both cases, though the time warping case presents additional mathematical difficulties. The modulation case was presented in a discrete, finite-dimensional setting, but can also be developed in the continuous time setting, following the lines of what was done for time warping.

The presented numerical simulations show the reliability of the proposed approach, and results on real world signals, though still preliminary, are quite encouraging. Further tests will involve validation on a database of real sounds, and may require perceptive tests. Among the possible improvements of our approach, a more thorough study of the optimization algorithms (for which we have chosen to stick to a simple and robust approach) will be a next goal.

\medskip
Let us stress that the motivation of this work was mainly in the field of audio signal processing, namely problems where an underlying transformation (related to a clock change, as is the case for sounds produced by rotating engines with variable rotation speed) is to be estimated. Nevertheless, the problem is fairly general and can be transposed to different contexts, such as the shape from texture problem in image processing (see~\cite{Clerc02texture}).

Let us also notice that the models introduced here form a new family of models for linear systems (or operators) represented directly in time-frequency or time scale space. Contrary to multiplier models (see for example~\cite{Feichtinger04approximation,Dorfler10time} that are defined by pointwise multiplications in the representation domains, and that can be estimated (see for example~\cite{Olivero13class}), these transformation models implement translations in that space. The proposed approach may therefore represent an interesting starting point for system estimation in the case of such categories of systems.

\section*{Acknowledgments}

This work was supported by {\em Agence Nationale de la Recherche} (ANR), in the framework of the Metason project (ANR-10-CORD-010).
H. Omer's is supported by {\em Agence Nationale de la Recherche} (same project).

\section*{References}
\bibliographystyle{IEEEtran}
\bibliography{OT}

\begin{thebibliography}{10}
\providecommand{\url}[1]{#1}
\csname url@samestyle\endcsname
\providecommand{\newblock}{\relax}
\providecommand{\bibinfo}[2]{#2}
\providecommand{\BIBentrySTDinterwordspacing}{\spaceskip=0pt\relax}
\providecommand{\BIBentryALTinterwordstretchfactor}{4}
\providecommand{\BIBentryALTinterwordspacing}{\spaceskip=\fontdimen2\font plus
\BIBentryALTinterwordstretchfactor\fontdimen3\font minus
  \fontdimen4\font\relax}
\providecommand{\BIBforeignlanguage}[2]{{%
\expandafter\ifx\csname l@#1\endcsname\relax
\typeout{** WARNING: IEEEtran.bst: No hyphenation pattern has been}%
\typeout{** loaded for the language `#1'. Using the pattern for}%
\typeout{** the default language instead.}%
\else
\language=\csname l@#1\endcsname
\fi
#2}}
\providecommand{\BIBdecl}{\relax}
\BIBdecl

\bibitem{Flandrin99time}
P.~Flandrin, \emph{Time-Frequency/Time-Scale Analysis}.\hskip 1em plus 0.5em
  minus 0.4em\relax Academic Press, 1999.

\bibitem{Huang05hilbert}
N.~E. Huang and S.~S.~P. Shen, \emph{Hilbert-Huang Transform and its
  Applications}.\hskip 1em plus 0.5em minus 0.4em\relax World Scientific, 2005.

\bibitem{Auger13time}
F.~Auger, P.~Flandrin, Y.-T. Lin, S.~McLaughlin, S.~Meignen, T.~Oberlin, and
  H.-T. Wu, ``Time-frequency reassignment and synchrosqueezing: An overview,''
  \emph{IEEE Signal Processing Magazine}, vol.~30, no.~6, pp. 32--41, 2013.

\bibitem{Wu13instantaneous}
H.-T. Wu, ``Instantaneous frequency and wave shape function (i),''
  \emph{Applied and Computational Harmonic Analysis}, vol.~35, pp. 181--199,
  2013.

\bibitem{Malik97computing}
J.~Malik and R.~Rosenholtz, ``Computing local surface orientation and shape
  from texture for curved surfaces,'' \emph{International Journal of Computer
  Vision}, vol.~23, no.~2, pp. 149--168, 1997.

\bibitem{Clerc02texture}
M.~Clerc and S.~Mallat, ``The texture gradient equation for recovering shape
  from texture,'' \emph{IEEE Transactions on Pattern Analysis and Machine
  Intelligence}, vol.~24, no.~4, pp. 536--549, 2002.

\bibitem{Clerc03estimating}
------, ``Estimating deformations of stationary processes,'' \emph{Annals of
  Statistics}, vol.~31, no.~6, pp. 1772--1821, 2003.

\bibitem{Omer13estimation}
\BIBentryALTinterwordspacing
H.~Omer and B.~Torr\'esani, ``Estimation of frequency modulations on wideband
  signals; applications to audio signal analysis,'' in \emph{Proceedings of the
  10th International Conference on Sampling Theory and Applications (SampTA)},
  G.~Pfander, Ed.\hskip 1em plus 0.5em minus 0.4em\relax Eurasip Open Library,
  2013, pp. 29--32. [Online]. Available:
  \url{http://hal.archives-ouvertes.fr/hal-00822186}
\BIBentrySTDinterwordspacing

\bibitem{Picinbono94circularity}
B.~Picinbono, ``On circularity,'' \emph{IEEE Transactions on Signal
  Processing}, vol.~42, no.~12, pp. 3473--3482, 1994.

\bibitem{Unser13sparse}
M.~Unser and P.~Tafti, \emph{Sparse stochastic processes}.\hskip 1em plus 0.5em
  minus 0.4em\relax Cambridge University Press, 2013.

\bibitem{Soendergaard07finite}
P.~Soendergaard, ``Finite discrete {G}abor analysis,'' Ph.D. dissertation,
  Institut for Matematik – Denmark Technical University, 2007.

\bibitem{Carmona98practical}
R.~Carmona, W.~L. Hwang, and B.~Torr\'esani, \emph{Practical time-frequency
  analysis: Gabor and Wavelet Transforms With an Implementation in S}, C.~K.
  Chui, Ed.\hskip 1em plus 0.5em minus 0.4em\relax Academic Press, 1998.

\bibitem{Grochenig01foundations}
K.~Gr{\"o}chenig, \emph{Foundations of time-frequency analysis}, ser. Applied
  and Numerical Harmonic Analysis.\hskip 1em plus 0.5em minus 0.4em\relax
  Boston, MA: Birkh\"auser Inc., 2001.

\bibitem{ltfatnote011}
P.~L. S{\o}ndergaard, ``{Efficient Algorithms for the Discrete Gabor Transform
  with a long FIR window},'' \emph{J.\ Fourier Anal.\ Appl.}, vol.~18, no.~3,
  pp. 456--470, 2012.

\bibitem{Mallat89theory}
S.~Mallat, ``A theory for multiresolution signal decomposition: The wavelet
  representation.'' \emph{IEEE Trans. Pat. Anal. Mach. Intell.}, vol.~11, pp.
  674--693, 1989.

\bibitem{Omer15Modeles}
H.~Omer, ``Mod\`eles de d\'eformation de processus stochastiques
  g\'en\'eralis\'es. application \`a l'estimation des non stationnarit\'es dans
  les signaux audio.'' Ph.D. dissertation, Aix-Marseille Universit\'e, 2015, in
  preparation.

\bibitem{Soendergaard12linear}
P.~Soendergaard, B.~Torr\'esani, and P.~Balazs, ``The linear time frequency
  analysis toolbox,'' \emph{International Journal of Wavelets and
  Multiresolution Information Processing}, vol.~10, no.~4, pp. 1\,250\,032--1
  -- 1\,250\,032--27, 2012.

\bibitem{Feichtinger04approximation}
H.~Feichtinger, M.~Hampejs, and G.~Kracher, ``Approximation of matrices by
  gabor multipliers,'' \emph{IEEE Signal Processing Letters}, vol.~11, no.~11,
  pp. 883 -- 886, 2004.

\bibitem{Dorfler10time}
M.~D\"orfler and B.~Torr\'esani, ``On the time-frequency representation of
  operators and generalized {G}abor multiplier approximations,'' \emph{Journal
  of Fourier Analysis and Applications}, vol.~16, pp. 261--293, 2010.

\bibitem{Olivero13class}
A.~Olivero, B.~Torr\'esani, and R.~Kronland-Martinet, ``A class of algorithms
  for time-frequency multiplier estimation,'' \emph{IEEE Trans. Audio, Speech
  and Language Processing}, vol.~21, no.~8, pp. 1550 -- 1559, 2013.

\end{thebibliography}
\end{document}